\pdfoutput=1
\documentclass[prl,twocolumn,reprint,superscriptaddress]{revtex4}

\usepackage[T1]{fontenc}

\usepackage{amssymb}
\usepackage{amsmath}
\usepackage{amsfonts}
\usepackage{graphicx}
\usepackage{subfigure}
\usepackage{dcolumn}
\usepackage{bm}
\usepackage{empheq}
\usepackage{natbib}
\usepackage{color,soul}
\usepackage{hyperref}

\begin{document}
\title{A machine-learning solver for modified diffusion equations}

\author{Qianshi Wei}
\affiliation{Department of Physics and Astronomy, University of Waterloo, Waterloo N2L 3G1, Canada}

\author{Ying Jiang}
\affiliation{
School of Chemistry, Beihang University, Beijing 100191, China}

\author{Jeff Z. Y. Chen\footnote{jeffchen@uwaterloo.ca}}
\affiliation{Department of Physics and Astronomy, University of Waterloo, Waterloo N2L 3G1, Canada}

\begin{abstract}{
A feed-forward neural network has a remarkable property which allows the network itself to be a universal approximator for any functions.
Here we present a universal, machine-learning based solver for multi-variable partial differential equations.
The algorithm approximates the target functions by neural networks and adjusts the network parameters to approach the desirable solutions.
The idea can be easily adopted for dealing with multi-variable, coupled integrodifferential equations, such as those in the self-consistent field theory for predicting polymer microphase-separated structures.
}\end{abstract}

\maketitle

 {\it Introduction.--}
Incorporation of machine-learning techniques \cite{james2013introduction, rasmussen2004gaussian, bishop2006pattern, michalski2013machine, Goodfellow-et-al-2016} into computational physics to tackle physical problems has dramatically changed the classical approaches in physics.
Supervised and unsupervised learning methods,
with their unsurpassed capability for practical applications such as image and voice recognitions, have found themselves a new playground in, for example, condensed matter physics.
Recent work has used machine-learning techniques to classify, manipulate, or even create
the big data produced for the structural and dynamic information of
various modeled systems \cite{carrasquilla2016machine,PhysRevLett.118.216401,van2017learning,PhysRevE.95.032504,PhysRevB.94.195105,PhysRevB.95.035105, PhysRevB.95.041101,PhysRevB.94.165134,Carleo602,PhysRevLett.108.253002, PhysRevX.7.031038, PhysRevE.97.013306, PhysRevLett.114.108001, sharp2018machine, PhysRevLett.120.176401}.

In this Letter we explore the usage of another fundamental property of neural networks, to solve partial differential equations (PDEs) used to describe physical systems by designing an unsupervised, universal machine-learning solver.
A common procedure is formulated regardless of the type of differential equations and the number of auxiliary conditions (initial conditions, boundary conditions, constraints, etc.).  We demonstrate the power of the procedure by solving modified diffusion equations in both high-dimensional and  %
%
complicated forms (e.g., coupled integrodifferential equations). The latter is encountered in predicting polymer microphase-separated structures, formulated from the self-consistent field theory   \cite{fredrickson2006equilibrium, doi:10.1063/1.430517, APOL:ACTP010460501, VILGIS2000167, 0953-8984-14-2-201, doi:10.1021/ma011515t, Muller2005Incorporating, hamley2004developments, Shi2015}.

The idea is to exploit a basic property of an artificial, feed-forward neural network (FNN), known as the universal approximation theorem. It states that any continuous functions can be effectively represented by FNNs, provided that adequate neuron nodes are used \cite{HORNIK1989359, Cybenko1989, HORNIK1991251}. The input nodes are simply the variables of the functions and the output nodes are functions themselves. The variety of functions are represented by the FNN parameters such as the weights and biases of the sigmoid functions that connect the neuron nodes \cite{haykin2004comprehensive}. If we can tune these FNN parameters to represent functions that satisfy partial differential equations (PDEs) and their auxiliary conditions, then we find a solution \cite{lagaris1998artificial}. The tuning is achieved by minimizing a cost function which embeds the targeted differential equations and the auxiliary conditions as squared modulus.

The computational concept is different from any traditional algorithms used to solve PDEs. In the latter case, the functions to be determined
are usually represented in some numerical form, by direct discretization or series-expansion on spectral bases; a traditional PDE solver adjusts these numerical values to satisfy the PDEs. Here, using FNNs, we adopt a different philosophy. 
The calculated functions are analytically represented by a universal network form, but with specific parameters determined through machine learning. In a sense, FNNs do not learn from the existing solutions of PDEs; they do, on the other hand, learn how to adjust themselves to satisfy the formal expression of PDEs. All complications involving stability analysis of a finite-difference method, for instance, are no longer the concern.

Reference~\cite{lagaris1998artificial} originally proposed that FNNs can be used to solve ordinary differential equations (ODEs) and its physics applications include finding plasma equilibrium state \cite{PhysRevLett.75.3594} and the time evolution of an $N$-body problem \cite{PhysRevLett.86.4741}. FNNs solvers were also proposed for low-dimensional PDEs \cite{lagaris2000neural, aarts2001neural} and their numerical accuracy in relationship with the network structures was recently discussed in Ref.~\cite{raissi2017physics}. The potential of using FNNs in high-dimensional problems is noted by \citeauthor{han2017overcoming} and \citeauthor{E2017} very recently \cite{han2017overcoming, E2017}. Note that independent sets of weighs and biases for different time frames were suggested in Refs. \cite{raissi2017physics,han2017overcoming,E2017}. In comparison, we document and analyze the solver's capability to handel high-dimensional PDEs, assigning the same weights and biases.

\begin{figure*}[!t]
\includegraphics[width=1.8\columnwidth]{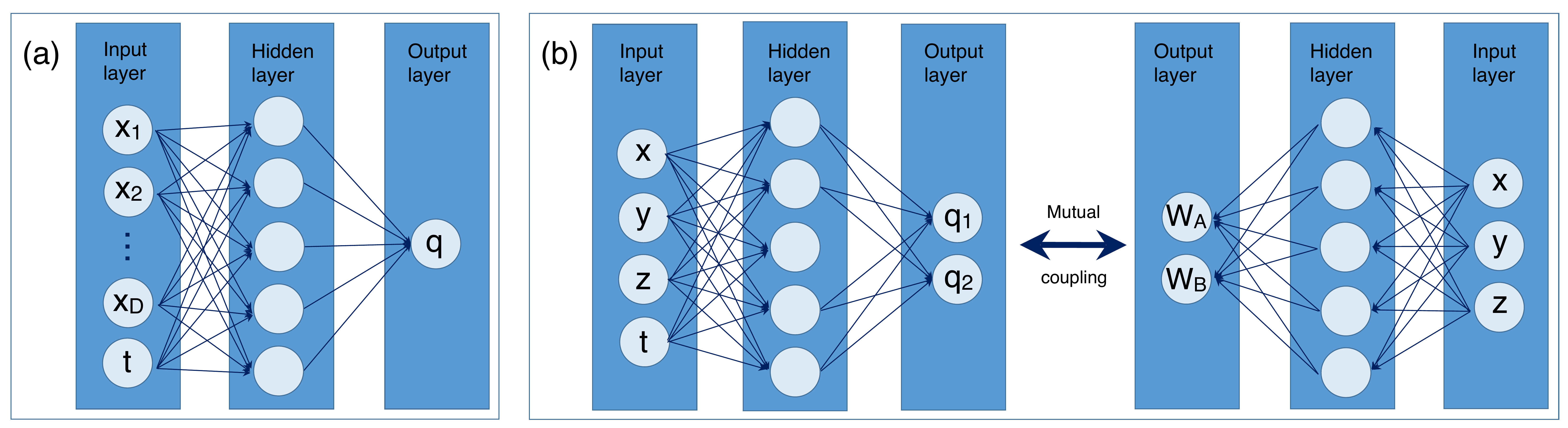}
\caption{\label{FIG0}Two examples of physical problems solved here: (a) a simple diffusion equation where $q({\bf x};t)$ is the density of the diffusing material in an external field at location ${\bf x}$ and time $t$ and (b) complicated, coupled modified diffusion equations where $q_1(x,y,z;t)$ and
$q_2(x,y,z;t)$ are the complementary reduced Green's functions for a real AB-diblock copolymer self-assembly problem, which couple to the self-consistent fields $W_A(x,y,z)$ and
$W_B(x,y,z)$.
In both examples, the functions to be found are represented by feed-forward neutral networks.
The circles represent neuron nodes, where the input layer consists of nodes that have variables as input and the output layer are simply the functions to be determined. The connections between the input and hidden layers are assumed to be sigmoid functions and the connections between the hidden and output layers are assumed to be linear with adjustable coefficients. 
}\label{FIG0}
\end{figure*}


{\it Main procedure.--} Consider well specified, coupled PDEs for functions $q_1({\bf r})$, $q_2({\bf r})$,... where the vector $\bf r$ generally represents multi-dimensional variables, and could be a combination of, for example, space and time variables.  
Generally, PDEs are
\begin{equation}\label{Dq0}
\hat{D}_1[q_1({\bf r}),q_2({\bf r}) ...]=0,  \hat{D}_2[q_1({\bf r}),q_2({\bf r}) ...]=0,...
\end{equation}
The differential operators, $\hat D_1$ and $\hat D_2$, act on the functions.
The problem is augmented by typical ``boundary conditions'' (or initial conditions if time variables are involved). For example, at boundaries ``1'', ``2'', etc.,
\begin{equation}\label{Ini0}
\hat{B}_1[q_1({\bf r}),q_2({\bf r}) ...]=0,  \hat{B}_2[q_1({\bf r}),q_2({\bf r}) ...]=0, ...
\end{equation}
{In addition, there could be constraints that govern these quantities, which are represented by
\begin{equation}\label{Con0}
\hat{C}_1[q_1({\bf r}),q_2({\bf r}) ...]=0,  \hat{C}_2[q_1({\bf r}),q_2({\bf r}) ...]=0, ...
\end{equation}
}
For abbreviation, the left hand sides are denoted as ${\hat D}_1({\bf r})$, ${\hat D}_2({\bf r})$, ${\hat B}_1({\bf r})$, ${\hat B}_2({\bf r})$, {${\hat C}_1({\bf r})$, ${\hat C}_2({\bf r})$}, etc.

In Fig. \ref{FIG0} we schematically illustrate FNN examples used in this work. At the initial stage, the parameters used in FNN are specified randomly or according to previous experience,  and hence in general the functions $q_1({\bf r}) $, $q_2({\bf r})$, ..., calculated from the FNNs are far from the desirable solutions. We design a cost function as
{\begin{equation}\label{Cost}
\begin{split}
J &= {\alpha_1\over 2} \left< \left\vert \hat{D}_1({\bf r}) \right\vert^{2} \right>
   +{\alpha_2\over 2} \left< \left\vert \hat{D}_2({\bf r}) \right\vert^{2}\right>+...
   \\
&+{\beta_1\over 2} \left< \left\vert \hat{B}_1({\bf r}) \right\vert^{2}\right>
   +{\beta_2\over 2} \left< \left\vert \hat{B}_2({\bf r}) \right\vert^{2}\right>+... \\
&+{\gamma_1\over 2} \left< \left\vert \hat{C}_1({\bf r}) \right\vert^{2}\right>
   +{\gamma_2\over 2} \left< \left\vert \hat{C}_2({\bf r}) \right\vert^{2}\right>+...
\end{split}
\end{equation}
}where
$\left <...\right >$ is the algebraic average of the quantity within, sampled at
a set of {randomly selected points} in the $\bf r$ domain. Upon the minimization of $J$ as a function of FNN parameters to reach $J=0$, the search finds an approximation of the represented functions $q_1$, $q_2$, ...
The coefficients, $\alpha_1$, $\alpha_2$,
$\beta_1$, $\beta_2$, {$\gamma_3$, $\gamma_4$,} ... are penalty coefficients that can be fixed or adjusted. 


A set of coordinates $\bf r$ (with values randomly selected from  its domain of interest), instead of the greyscale pixel values in pattern recognition \cite{bishop2006pattern},  are used in the input layer as a single ``sample''. Through FNN, the output data produces a guess of the functions to be studied. Within an epoch, many such randomly selected samples are used to produce guesses of the functions at different points in the domain.
During the training session, $J$
is used to minimize the mean-square averages of the left-hand sides of Eqs. \eqref{Dq0}, \eqref{Ini0} and \eqref{Con0}, which are calculated according to the outputting, machine-guessed functions.
One epoch of minimization is then performed.
No numerical solutions obtained from any other methods are used in this procedure.


{\it Diffusion equation.--}
Taking the diffusion equation for illustration, in $D$ spatial dimensions we write
\begin{equation}\label{DED}
{\hat D} q({\bf x};t) = \left [{\partial \over \partial t} - {1\over 6} \sum_{n=1}^{D} {\partial^{2} \over \partial x_{n}^2} + W({\bf x})  \right ] q({\bf x};t) =0.
\end{equation}
where ${\bf x} =(x_{1}, x_{2},\ldots, x_{D})$ is a $D$-dimensional vector. Any traditional numerical method to solve this requires the computation to determine at least $K^{D+1}$ representative data points.
For example,
the finite difference method directly divides the $D$-dimensional space into representative nodes,
where on average $K$ nodes for each $x_n$ are needed.
Taking an under-estimate that a tradition algorithm is linear in $K^{D+1}$ to achieve a solution of precision $\epsilon$, in high-$D$ this amounts to exponential growth in computational time and storage resource \cite{Papadimitriou:2003:CC:1074100.1074233}. Most real algorithms \cite{press2007numerical}, of course, are more expensive than $K^{D+1}$. This problem is known as the curse of dimensionality \cite{bellman2013dynamic}.

\begin{figure}[!t]
         \includegraphics[width=0.9\columnwidth]{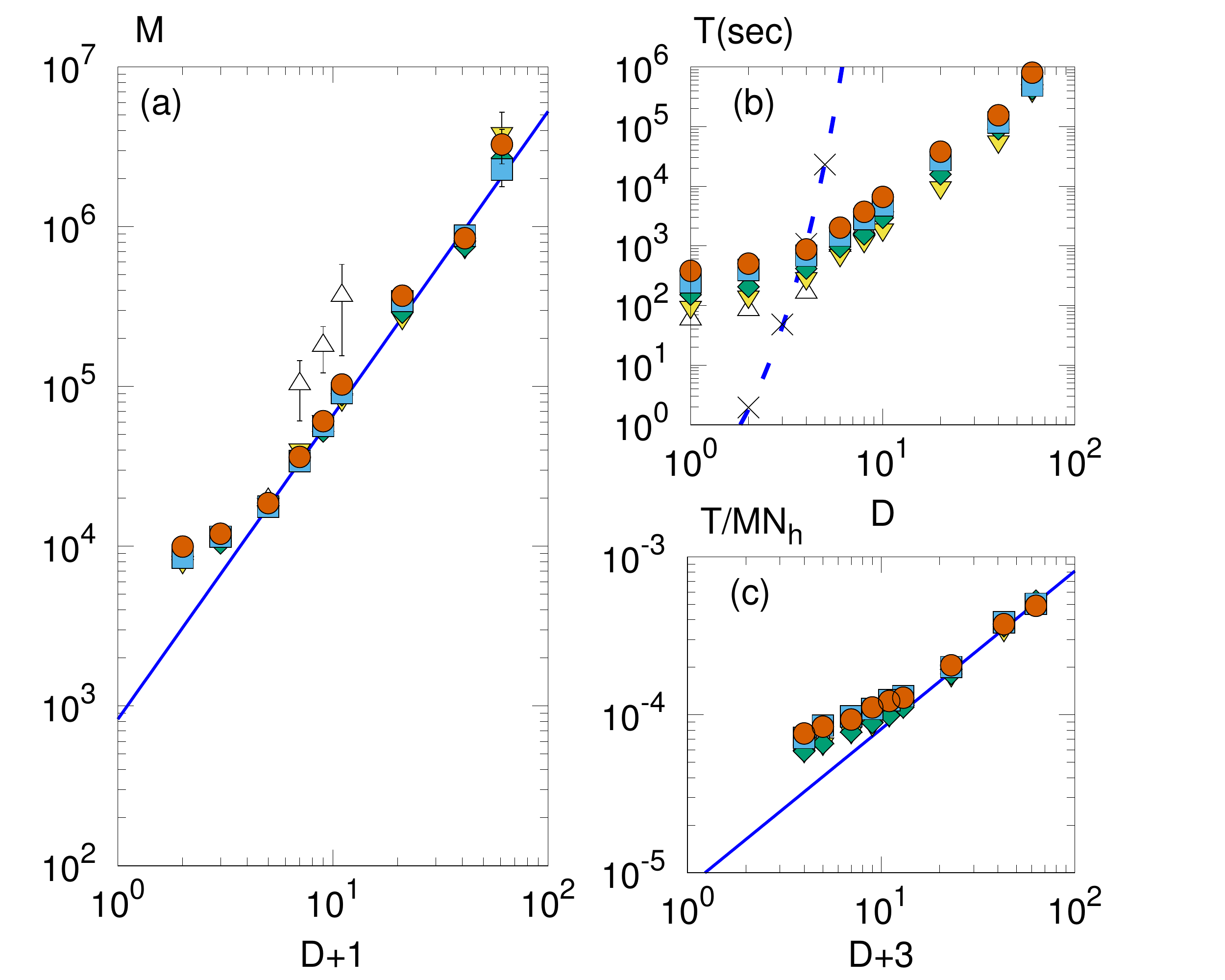}
        \caption{\label{eD}
        Log-log plots of (a) the maximum epochs $(M)$, and (b) the total computational time $(T)$  that the universal solver takes to reach the  error tolerate level $\epsilon=10^{-3}$, as functions of $D$, the number of spatial variables in  a high-dimensional diffusion equation, Eq. \eqref{DED}; (c) Log-log plot of the ratio $T/MN_h$ as a function of $D+3$. The error bars, estimated from 10 independent runs, are smaller than the plotted symbols, except for those explicitly shown.
        Up triangle, down triangle, diamonds, squares, and circles represent the results produced from FNNs that contain
        $N_h=100,200, 300, 400$, and $500$ hidden nodes, respectively.
         The solid blue lines indicate the power laws on which the data points collapse.
         The dashed blue curve illustrates the exponential dependence of the computational time required by the traditional Euler's solver, where the crosses are the actual tested time at $D+1=3,4,5$, and $6$ \cite {SM}.
	}\label{CD}
\end{figure}

Our universal solver takes the a different approach.
The number of nodes in the hidden layer, $N_h$, and the maximum epoch loops, $M$, required in a learning process to achieve a pre-specified precision $\epsilon$,
directly determine  the computational complexity. To understand the dependence of $M$ on $D$, as an example, we numerically solve
Eq.~\eqref{DED} in a specific potential field $W({\bf x} ) =(1/2) \sum_{n=1}^D x_n^2$, incorporating an initial condition ${\hat B}_1 q({\bf x}; 0) = q({\bf x}; 0) -1=0$, for selected $D$ up to $10^2$. Other technical parameters include: for every training epoch the selection of $S=500$ sample points [each $(D+1)$-dimensional] in the $D+1$ dimensional space spanned by 
$({\bf x}; t)$,
pre-specified error tolerance $\epsilon=10^{-3}$ for $J$, and the placement of $20\%$ of the sampling points at $t=0$ to handel the initial condition.
To collect adequate statistics, for a given $D$ we conducted 10 separate learning runs, each starting from  {a random selection of the FNN parameters from normal distribution of mean $0$ and variance $0.1$}. A data point in Fig. \ref{CD} is an average from these 10 runs.

To explore the complexity of the problem, we select different $N_h$ for various $D$.
%
A striking feature of Fig. \ref{CD}(a) is that 
$M$ follows a linear behavior at large $D$ on a double-logarithmic plot, with a slope $\nu \approx 1.9$, 
\begin{equation}\label{power}
M \propto D^\nu.
\end{equation}
Although we are unable to analytically deduce this dependence, the numerical evidence indicates a rather optimistic scaling property for the required computational loops with a common exponent $\nu$, for a large $D$ up to a limit fixed by $N_h$.

We now estimate the computational resource required to solve a problem.
The FNN structure is described in the Supplemental Material.
The total number of FNN parameters $P=(D+1)\times N_h+ N_h+N_h\times 1$, where the three terms are for the number of $w$-parameters between the input and hidden layers, the number of $b$-parameters on the hidden layer, and the number of $v$-parameters between the hidden and output layers, respectively. An immediate advantage is computational storage. Our solver memorizes $P=(D+3) N_h $ parameters instead of approximately $K^{D+1}$ representative nodes.

Another main concern is the computational time. On each epoch pass, $P$ parameters need to be updated. The computational time of the back-propagation method \cite{bishop2006pattern} linearly depends on $P$, and hence the total computational time $T$ is 
\begin{equation}\label{T}
T \propto MP = M(D+3) N_h.
\end{equation}
This is a surprisingly pleasant power-law scaling
in contrast with the exponential law $K^{D+1}$ illustrated in plot-(b)
anticipated from a traditional approach. For comparison, at $D=60$, {the projected computational time of Euler's method takes approximately $7\times 10^{72}$ years on a moderate $K=24$ divisions of each variable }\cite{SM}, {whereas the machine solver presented here takes approximately 12 days. }

{\it Mesoscopic structures in diblock copolymers. --}
Next we solve a rather complicated integrodifferential equation set for a classical computational problem in polymer physics \cite{SM}. {The goal here is to test the capability of the machine-learning solver to deal with a rather mathematically involved, classical theory to describe a real-world problem.}

\begin{figure}[!t]
        \centering
	\includegraphics[width=1.01\columnwidth]{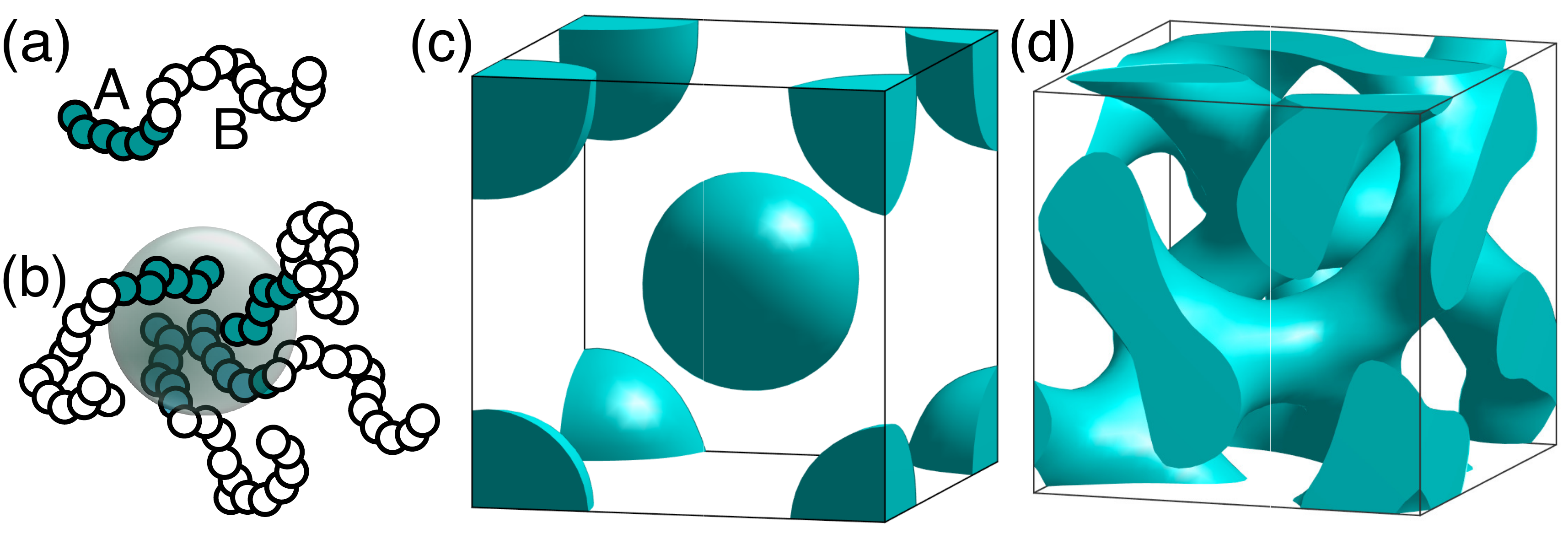}
        \caption{\label{BCC3Dsmall}
        Solving the self-consistent field theory for the microphase structures of diblock copolymers. Plot (a) illustrates a single polymer chain where a covalent bond links A and B blocks together. Plot (b) shows the cross section of an A-rich spherical domain in a B-rich background. Plots (c) and (d) are our three-dimensional numerical solutions of the monomer-fraction profiles from the theory, which have body-centered cubic and gyroid structures, respectively. The solutions are obtained from a machine-learning algorithm that incorporates representations of functions conceptually shown in Fig. \ref{FIG0}(b).
        For illustration purpose, we plot all A-rich regions with the same green color. }\label{Poly}
\end{figure}

{Here is a short summary of the physics we wish to tackle: structural prediction of a densely packed, molecular system known as diblock copolymer melt. Each long-chain molecule, as shown in Fig.} \ref{Poly}(a), {contains two incompatible blocks, consisting of A- and B-type molecular units, respectively represented by green and white. The physical question is: what is the crystallographic structures when many of these copolymers are densely packed in a finite volume to form possible periodic structures? Examples of the overall structures are shown in Figs.} \ref{Poly} {(b) and (c), where the green units stay in the green domain and white in white. } \cite{Bates434}.

The well-developed self-consistent field theory {(SCFT)} is a useful tool for structural predictions in these systems \cite{1994FaDi_Bates,matsen1996MM,2010Science_Bates,2014PNAS_Lee,2017Science_Bates,2018Kim847}.
The complicated mathematical structure of the SCFT is listed in the Supplemental Material \cite{SM}. 
Four basic, unknown functions $q_1 ({\bf r})$, $q_2({\bf r})$, $W_A({\bf x})$, and $W_B({\bf x})$, must be found numerically, under a given molecular architecture. Both functions, $q_1 ({\bf r})$, $q_2({\bf r})$, satisfy modified diffusion equations where $W_A({\bf x})$, and $W_B({\bf x})$ are the external-field components. The functions $W_A({\bf x})$ and $W_B({\bf x})$ are dependent on $q_1 ({\bf r})$ and $q_2({\bf r})$ by integrations. In addition, there are boundary conditions and other constraints one needs to deal with.
In a traditional approach, multiple iterations are needed to achieve the self-consistency of the solution set. The main idea there is to propose a guess for the external fields $W_A({\bf x})$ and $W_B({\bf x})$, which are used in the diffusion-like equations governing the  propagator functions $q_1 ({\bf r})$ and $q_2({\bf r})$. Integrating over the $t$ variable step by step, one  thus obtains the solutions for $q_1 ({\bf r})$ and $q_2({\bf r})$. The external fields $W_A({\bf x})$ and $W_B({\bf x})$ are then updated according to these solutions and a new iteration step starts. Self-consistency is obtained after multiple iteration loops, at which point the $W$ fields converge. Typical classical algorithms are well-documented in a book \cite{fredrickson2006equilibrium}.

A completely different philosophy is adopted here, in order to implement the machine-learning solver. The four functions are presented by two FNNs, conceptually shown in Fig. \ref{FIG0}(b). The learning is done by looping through epochs.
At every epoch, the FNNs learn the updated profiles of these functions simultaneously by renewing the FNN parameters, according to the minimization requirement of the cost function. 
{The cost function ($J$) itself is the sum of terms that are targeted at solving the modified diffusion equations ($J_D$), that effectively deal with the boundary conditions for given $t$ and given $\bf x$ ($J_{\rm ic}$ and $J_{\rm bc}$), and that couple $W_A$, $W_B$ with $q_1$, $q_2$ nonlinearly in an integral form ($J_C$).}
There is no need to integrate the differential equations over $t$-domain step by step, because $t$ is now treated at an equal footing as $\bf x$. 
Most importantly, the iteration loop that updates $W_A({\bf x})$ and $W_B({\bf x})$ 
is now eliminated. The self-consistency is directly enforced through the non-linear coupling of the four functions.

Starting from a random choice of the FNN parameters, our machine-learning solver
reproduced all known three-dimensional classical structures, such as those presented in Refs. \cite{1994FaDi_Bates,matsen1996MM,2010Science_Bates,2014PNAS_Lee,2017Science_Bates,2018Kim847}, within their own stability parameter region.
Two of which are presented in Fig. \ref{Poly}(c) and (d) for visual presentation.
There is an excellent agreement between our and previous solutions.

Searching for stable and meta-stable conformations of a polymeric system is a long enduring and crucial topic in soft matter physic in large part due to their rich and complex self-assembly behavior \cite{doi:10.1146/annurev.pc.41.100190.002521, Whitesides2418, POLB:POLB21334, PhysRevLett.96.138306, wzg201750th}. The heart of a theoretical approach is to solve SCFT equations in order to make structural prediction. Here we wish the machine-learning solver can overcome the main hurdle encountered in a conventional method --- the stability of a proposed algorithm.

{\it Summary. --}
Taking advantage of the universal approximation theorem, we present a machine-learning procedure designed to solve partial differential equations.
We started by introducing a fundamental diffusion problem and then continued by tackling a complicated integrodifferential equation set produced from the modern polymer theory.

Our solver avoids the potential pitfalls typically seen in a traditional numerical approach.
The approximations for the derivative operators in the partial differential equations are no longer needed and all required information are expressed by analytic expressions, through the representing FNNs. The approximation made in a traditional method highly influences the stability of a typical algorithm such that the stability of a computational algorithm usually becomes the main concern. Here, this difficulty is avoided by turning the solution-finder problem into a machine-learning problem.
{The machine-learning solution usually converges in the sampled variable space multilaterally } \cite{SM2}.
The solver is an unsupervised procedure that
requires no {\emph {a priori}} information of the solution and accommodates boundary conditions and constraints systematically.

The information storage in a network of $P$ parameters for a $D$-dimensional function, however, has an upper limit $D \lesssim P$. Hence the power law in \eqref{power} cannot remain valid in an asymptotically large $D$ for a fixed $N_h$.
The traditional curse of dimensionality (represented by the exponential law $N^D$)
is partially broken within the limit set by $N_h$. Modern computer and computation technologies can boost $N_h$ to a fairly large number. Thus, we expect that this universal solver is particularly useful for solving physical problems containing many variables and coupled functions.

{We adopted randomly selected sampling points from a uniform distribution in the interested multi-variable domain, similar to a }{\emph{simple}}  {Monte Carlo method. Further improvements could include the weighted Monte Carlo method, either by a prescribed weight} (umbrella sampling \cite{umbrSample,frenkel2001understanding}), {or, perhaps by directly using the squared curvature of the approximated function produced from the previous epoch as the weight.}



The authors wish to acknowledge the financial support from the Natural Science and Engineering Council of Canada (NSERC), Beihang University, and the National Science Foundation of China (NSFC, 21574006, 21622401). 

\bibliography{ref}

\end{document}